# Different sensing mechanisms in single wire and mat carbon nanotubes chemical sensors

*P.L. Neumann[a,b], V.I. Obreczán[c], G. Dobrik[a], K. Kertész[a], E. Horváth[a], I.E. Lukács[a],*

*L.P. Biró[a] and Z.E. Horváth[a]*

[a] *MTA TTK MFA, Institute of Technical Physics and Materials Science, Research Centre for Natural Sciences, H-1525 Budapest P.O. Box 49, Hungary*

[b] *Budapest University of Technology and Economics, Department of Electronic Devices, Budapest, H-1521 P. O. Box 91, Hungary*

[c] *Budapest University of Technology and Economics, Solid State Physics Laboratory, Department of Physics, Budapest, H-1521 P. O. Box 91, Hungary*

*Corresponding author: P.L. Neumann, Neumann.Peter@ttk.mta.hu; addr.: Budapest, H-1525 P. O. Box 49, Hungary; tel.:+36-1-3922222/1157 ext.*

**Abstract**

Chemical sensing properties of single wire and mat form sensor structures fabricated from the same carbon nanotube (CNT) materials have been compared. Sensing properties of CNT sensors were evaluated upon electrical response in the presence of five vapours as acetone, acetic acid, ethanol, toluene, and water. Diverse behaviour of single wire CNT sensors was found, while the mat structures showed similar response for all the applied vapours. This indicates that the sensing mechanism of random CNT networks cannot be interpreted as a simple summation of the constituting individual CNT effects, but is associated to another robust phenomenon, localized presumably at CNT-CNT junctions, must be supposed.

**Keywords**

organic vapour detection, chemical sensor, capillary condensation, CNT-CNT interconnection

**Introduction**

The problem of chemical sensing has been an area of continuous interest: developing more selective, sensitive, cost-effective sensors for various chemicals is the endeavour of many industry and academia research groups. Carbon nanotubes (CNT) have many attractive properties making them suitable as sensor materials. CNT chemical sensors based on different principles have been demonstrated, including resistive, capacitive sensors, chemical field effect transistors, gas ionization sensors [1], thermoelectric response [2], or CNT-aided acoustic or optical sensors [3]. Besides, numerous studies of CNT chemical sensing mechanisms by different simulation methods have been published [4-7] mostly for relatively simple CNT/molecule systems, to find the correct theory of sensing principle.

Resistive CNT sensor implementations can be classified into two groups: one is the family of sensors based on a single wire when an individual CNT or single-wall CNT (SWCNT) bundle with metallic contacts on both ends is the active sensor material and the other where bulk densities are applied. This latter group involves the cases of aligned CNT arrangements prepared either by dielectrophoresis [8] or by low-pressure chemical vapour deposition [9] which are not discussed here, as well as the random networks prepared by filtering [10], printing [11], or suspension drop-drying [12]. If CNTs are not needed to be manipulated individually, preparation methods could be obviously much simpler. This justifies the spread of bulk CNT sensors, although single wire CNT sensors can reach much higher sensitivity. Most of the studies on the chemical sensing properties of CNTs report on the preparation of sensors made from CNT of one certain type and origin using either a single tube or a bulk structure [1].

The aim of the present work was to compare the chemical sensing properties of mat and single wire sensor structures made of the same CNT material in order to learn how the behaviour of individual CNTs determines the mat properties. Both, single- and multiwall CNT resistive sensors of different origin, were tested under the stream of five different vapours by measuring the electrical response. Both selected CNT materials showed significant sensitivity for the same vapours in a previous study [10].

**Experimental**

One of the applied CNT materials was a multiwall CNT (MWCNT) sample grown by catalytic chemical vapour deposition (CCVD) on Co-Fe/alumina catalyst [13] having a characteristic diameter of 10-20 nm and a typical length of 30-50 µm. The other CNT material was a SWCNT sample synthesized by the arc discharge method using Ni-Y catalyst [14], and then purified by refluxing in 3 M $HNO_3$ for 45 h. The sample contains SWCNT bundles with diameters ranging from 10 to 20 nm; besides, carbon-coated catalyst particles, empty graphitic shells, and a few MWCNT are present (Fig. 1). In our experiments, debundling of the SWCNT was not intended in order to avoid surfactant contamination. The percentage of –COOH groups in the SWCNT mat sample was found to be about 5 % by XPS study, using the decomposition of C 1s peak (not shown here).

The fabrication of single wire and mat CNT sensors needed substantially different processes. In case of single wire CNT sensors, irrespectively of the CNT type, we used Si/$SiO_2$ wafer substrates pre-patterned with Cr/Au marker and electrode/pad system. The CNT powders were dispersed in ethanol by sonication to obtain homogeneous [15] suspensions using a Cole Palmer CP750 ultrasonicator. CNT suspensions were spin-coated with 3000 rpm at room temperature over markered substrates. The concentration of the suspensions was approx. 10 pg/ml, which is an empirical density for reaching about one CNT over 40 $\mu m^2$ density. This was optimal for contact preparation by electron beam lithography (EBL). Veeco MultiMode8 atomic force microscope (AFM) was used in tapping mode (Fig. 2a) to locate CNTs or CNT bundles suitable for contacting, in order to avoid contaminations originating from scanning electron microscopy (SEM) investigation. The electric contacts were deposited by standard EBL lift-off technique to connect the single nanotubes to the measurement pads. The Cr (15 nm)/Au (35 nm) metal layers were evaporated through an exposed and developed 300 nm thickness poly-methyl methacrylate (PMMA) mask onto the surfaces.

The CNT mat sensors were prepared by the method described in the paper of Horvath et al. [10] where the mats of nanotubes were formed by filtering the suspensions through polycarbonate membrane filters with 400 nm hole size. Circular dots of the filtered CNT layers were formed with about 1.5 mm diameter and approximately uniform layer thickness. The nanotube amount in the layers was about 0.6 µg/mm$^2$. Finally, the middle part of the sensor elements was masked during evaporation of a gold layer with cca. 300 nm thickness, resulting in an uncovered zone with approximately 1 mm$^2$ active area between two gold covered stripes (Fig. 2b).

All the prepared sensor structures were tested by I(V) measurements before sensing experiments to sort out the defective ones.

For gas sensing experiments, the sensing system for vapour exposition was built up with two gas loops with a mechanical valve switching between them (Fig. 3). Pure $N_2$ flowed in the neutralization loop (Loop A) used in the initial and final state of the measurements for purging the sensor with a flow rate of 1700 ml/min. The nitrogen gas in the second measurement loop (Loop B) flowed through a bubbler, containing the liquid used as vapour source. The mixture of nitrogen and the studied vapour was led to the socket containing the sensor. The effects of acetone, acetic acid, ethanol, toluene and water were studied. The duration of each measurement cycle was 500 s with three stages: the first, reference stage took 100 s, the measurement stage 200 s and the final, purging stage 200 s, respectively. In the first stage (t= 0-100 s), the gas atmosphere was $N_2$, the average of resistance values measured during this period was used as reference. At t= 100 s, the position of the manual valve was changed to let the $N_2$ gas flowing in Loop B trough the bubbler containing the liquid of the measured compound. The sensor in this period was exposed to the mixture of $N_2$ and the saturated test vapour. The third stage started at t= 300 s when the valve was turned back again to purge the system with pure $N_2$ gas (Loop A).

The chemical sensing properties of the sensors were characterized by comparing the electrical resistance measured in pure nitrogen and nitrogen + vapour atmosphere. Two-point probe electrical transport measurements during each cycle were taken by a Keitley 2400 Sourcemeter with Süss probes at room temperature with sampling rate of 5 ms. The average of 100 measured data was read out from the instrument to decrease the noise, resulting in a plotted sampling frequency of 2 Hz. The set points of the electrical measurements were determined to keep the energy dissipation low enough for preventing samples from damaging. The set point of electric response measurements on gas exposure was approx. 1.5 mW and 10 nW in current generator mode in case of mat and single wire CNT sensors, respectively. Measurements with several energy stimuli from 3 nW to 2 mW using the same configuration were also taken in case of mat sensors to monitor the tentative set point dependence of the response. The response amplitudes were found to be not more than 30 % higher in cases of the lowest power compared to the highest one however the curve shapes were very similar for all the different set points (not shown here).

**Results and Discussion**

All the investigated sensor units were found to show metallic behaviour by I(V) measurements (not shown here). CNTs are known to be either metallic or semiconducting, depending on their chirality, the specific arrangement of the carbon atoms in the cylindrical wall [16]. The gap of the semiconducting CNTs is inversely proportional with the tube diameter [17]. In case of MWCNT, the outmost shell is dominating in the electric transport [18]. In our case, the diameter of the investigated MWCNTs falls typically between 10-20 nm. The band gap energy of semiconducting CNTs with such a large diameter are below 100 meV, so the number of thermally excited charge carriers at room temperature is high enough for metallic behaviour [19, 20]. SWCNTs in our experiments were applied in the form of bundles. Though the chirality of individual SWCNTs were not studied in our case, it is well known

[21] that usually about one-third of them is statistically metallic and the rest is semiconducting. SWCNT bundles contain nanotubes parallel with each other, if any of them is metallic, the combined behaviour is approximately metallic at room temperature. The diameter of the investigated SWCNT bundles was at least 10 nm, we can assume that they contain statistically at least 20 individual nanotubes. In this case, the probability that all of them are semiconducting is very low. According to the above considerations, the found metallic behaviour of the sensor structures corresponds to the expectations.

The relative resistance changes in the investigated MWCNT and SWCNT mat and single wire sensors are plotted for different vapour expositions as a function of time (Fig. 4). The response curves of the different vapours are displayed alternating in black and grey colours and shifted in y direction for better visibility. The plotted relative resistance ($R_{rel}$) is given by $R_{rel} = R_V / R_{N2}$, where $R_V$ is the resistance measured during vapour flow experiment, while $R_{N2}$ is the resistance measured in the initial stage during only pure $N_2$ exposition.

The average resistance of mat sensors was less than 1 kΩ, in agreement with literature results [18, 22-24]. The total resistance of the contacted single wire CNT sensors exceeded several tens of MΩ; thus, the injected current had to be kept lower than that of mat sensors, causing higher noise levels. An FFT low-pass parabolic filter with 0.33 Hz cut-off frequency was used to reduce the noise effect [25] with the same filtering parameters for all the single wire CNT sensor response curves. The SWCNT sensors, both mat and single wire ones, exhibited larger response amplitudes compared to MWCNT ones, the average maximum of the relative resistance curves was approximately 35 % in case of the SWCNT mat and 1 % in case of the MWCNT mat sensor, take note the different scales on Fig. 4. Finding the correct explanation of this difference needs more detailed study of the characteristics of conduction mechanism in the two corresponding samples which is far beyond the scope of the present paper. We assume that the main reason can be one of these: either the higher specific surface area or different functional group coverage of SWCNTs.

The differences in the charts of Fig. 4 show the diverse behaviour of the single wire CNT samples due to exposure to different vapours. The curves are heavily noise involved and therefore uncertain, and because of the low power level, the sign of the resistivity changes can still be clearly identified in some cases. Aside from noise effects, repeated measurements produced similar responses. Single wire MWCNT sensor responded unambiguous resistance decrease for acetone while no characteristic response above the noise level can be identified for the other vapours. The SWCNT sensor showed decreasing response for acetone and acetic acid, increasing response for ethanol and water but no characteristic change for toluene. As for the curve shapes, there is apparently no direct correspondence of the signal level to vapour concentration, but the quick changes in ambient composition trigger the changes in different sign, rate, and complexity, depending on the corresponding sensors and vapours.

Mat sensors, however, both SWCNT and MWCNT ones, showed similar behaviour for all the five applied vapours, namely the increase in the resistance when vapour exposition started, tending to a saturation value, decrease when vapour exposition finished, and tending to a new saturation value. The visible significant differences can be attributed to differences in transition times of these processes. Besides, a slow change in the baseline can be observed, especially in case of the MWCNT sensor, where the rate of this is comparable to the much smaller response amplitudes.

Reproducibility of the measured response curves was found to be within 3 % regarding the response amplitude maxima if the depletion of the previous vapour from the system was complete.

Surprisingly, the characteristics of response signals were completely different in many cases when single wire and mat sensors made of the same material were compared, in spite of the similar measurement conditions. In case of the SWCNT sensor, the signs of the changes were opposite for acetone and acetic acid, while in case of the MWCNT sensor for acetone. These observations suggest that the vapour-induced resistance changes in nanotube mats cannot be explained simply as the resultant of the individual nanotube resistance changes. From the response signals, we can assume that the dominant sensing mechanisms in case of resistive single wire and mat CNT sensors are different.

The conventional sensing principle of CNT sensors is based on the model that adsorption or desorption of a molecule makes changes in the CNT local electronic structure, donates or withdraws charge carriers to or from the electrical circuit, which can be detected as resistance change [26]. Structural defects and functional groups can serve as adsorption sites, this way the relative density of these imperfections can have significant effect on the sensor efficiency. Our observations in case of single wire CNT sensors can be in accordance with these considerations, regarding the probable differences in the defect structure and functional group coverage of the two investigated CNT types.

In case of mat structures, high number of CNTs is involved in the conduction and the current flows through many CNT-CNT junctions. The resistance in CNT networks is known to be dominated by the resistance of network junctions [27]. This does not necessarily mean that resistance change induced by the change in the chemical ambient is caused by the change in the junction resistances. However, considering that the junctions are not present in single wire sensors, the effect explaining the dissimilar behaviour of the two sensor types should primarily be found among the junction related ones.

The role of CNT-CNT junctions in chemical sensing can be evaluated by studying individual junctions. Here, we report on a preliminary chemical sensing experiment using an individual junction of two SWCNT bundles partially seized together longwise. We used the above-mentioned SWCNT material and the same contacting procedure as in case of single wire sensors detailed above. AFM image of the contacted structure is shown in Fig. 5a. The effect of water vapour was studied only, the structure was damaged during second vapour test. The response curve of the individual junction for water, as well as the response curve of a simple single wire SWCNT sensor prepared on the same chip and measured simultaneously with the junction is presented in Fig. 5b. The relative resistance increase found in case of the junction corresponds to the behaviour of the mat sensors, while the single wire sensor in this case showed resistance decrease. These findings do not prove but at least indicate the validity of the above idea on the importance of the junction resistance change in the chemical sensing mechanism of CNT mat sensors. Differences between individual structures can be significant as it is shown by the opposite response of the two single wire SWCNT sensors for water presented on Figs. 4 5, respectively.

An obvious possible reason of resistance change in a CNT network where the nanotubes are linked together mostly by weak Van der Waals forces is the change in junction gaps. Kumar et al. [28] suggested it as sensing mechanism in case of polymer-coated random MWCNT network sensors. In their case, the change in junction gaps was caused by the swelling of polymer layer covering the CNTs due to organic molecule's diffusion. Not covered by polymer, another effect of the chemical ambient change on the junction resistance must be supposed in our CNT mat sensors. A possible explanation is capillary condensation, the process when, at confined geometries, a liquid condensates into the pores of a medium from vapour phase at a pressure lower than the saturated vapour pressure at the same temperature [29]. The presence of a tiny liquid condensate between two CNTs can change the junction resistance directly by affecting the tunnelling probability or by displacing the nanotubes and modifying the junction gap. If the ambient vapour concentration rises, the conditions of capillary condensation can be fulfilled at more and more junctions. Perfect CNT is known to be hydrophobic [30], which rules out capillary condensation of water but the presence of structural defects, functional groups, can locally change its wetting properties. There are evidences that water wets, at least locally, CVD grown CNTs [31].

To judge if it has marginal or dominating impact on the sensor response needs a comprehensive theoretical analysis of capillary condensation at CNT junctions, including the study of junction geometry changes and conduction properties in the presence of a liquid condensate. Correct analysis of this problem is beyond the scope of the present paper.

**Conclusion**

Resistive sensors fabricated from individual MWCNT and single wire SWCNT bundles on the one hand and from mats of them on the other hand were used to compare the chemical sensing properties of single wire and bulk CNT structures. Single wire CNT sensors showed varied characteristics also in the sign of change, while in case of mats the exposure to the saturated vapour of five volatile test chemicals induced uniformly conductivity decrease of about 35 and 1 % for SWCNT and MWCNT sensors, respectively. The similar behaviour of mat sensors beyond the quantitative differences in response amplitude and transition times indicates that the sensing mechanism in case of mat structures is dominated by an effect which is not significant in case of single wire sensors. A preliminary measurement on an individual SWCNT junction is in agreement with this idea. A probable effect inducing conductivity changes in case of ambient vapour concentration variations is capillary condensation. Tiny liquid condensates present at the confined spaces of CNT-CNT junctions with volume varying with the vapour concentration can displace the nanotubes or change tunnelling probability effecting the conductivity variation in the whole network.

**Acknowledgments**

The authors are grateful to Atilla SULYOK for the XPS study of the SWCNT sample.

**Figures**

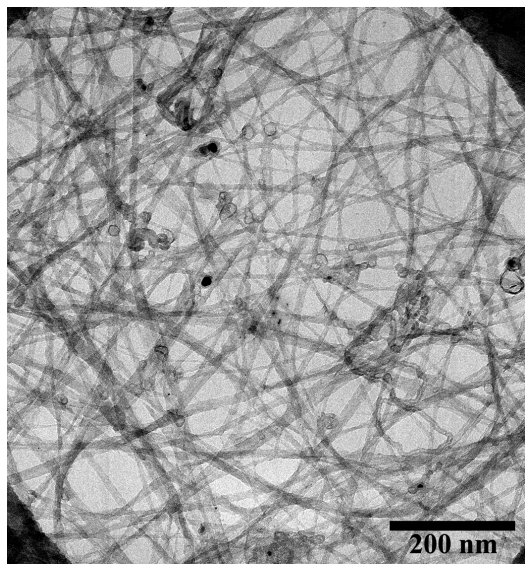

**Fig. 1** TEM image of purified SWCNT material used in the experiments.

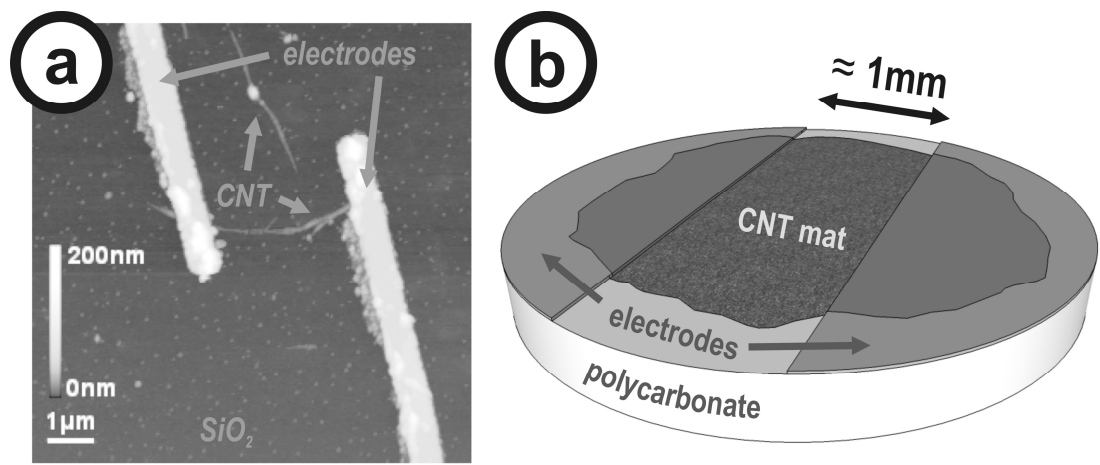

**Fig. 2**(a) AFM image of an individual SWCNT bundle contacted with Cr/Au electrodes on Si/SiO$_2$ substrate, (b) schematic drawing of CNT mat contacted with Au electrodes on polycarbonate substrate, used as sensor.

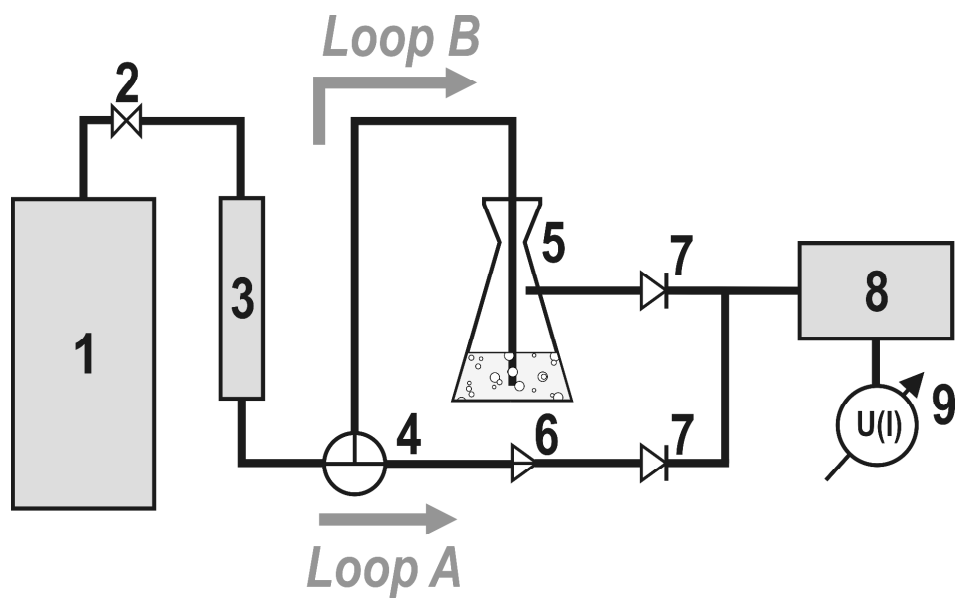

**Fig. 3** Scheme of the measuring system: (1) N$_2$ tank, (2) close/open valve, (3) flow metre, (4) two-way valve, (5) bubbler bottle, (6) restrictive, (7) one-way valves, (8) sample, and (9) transport measurement system.

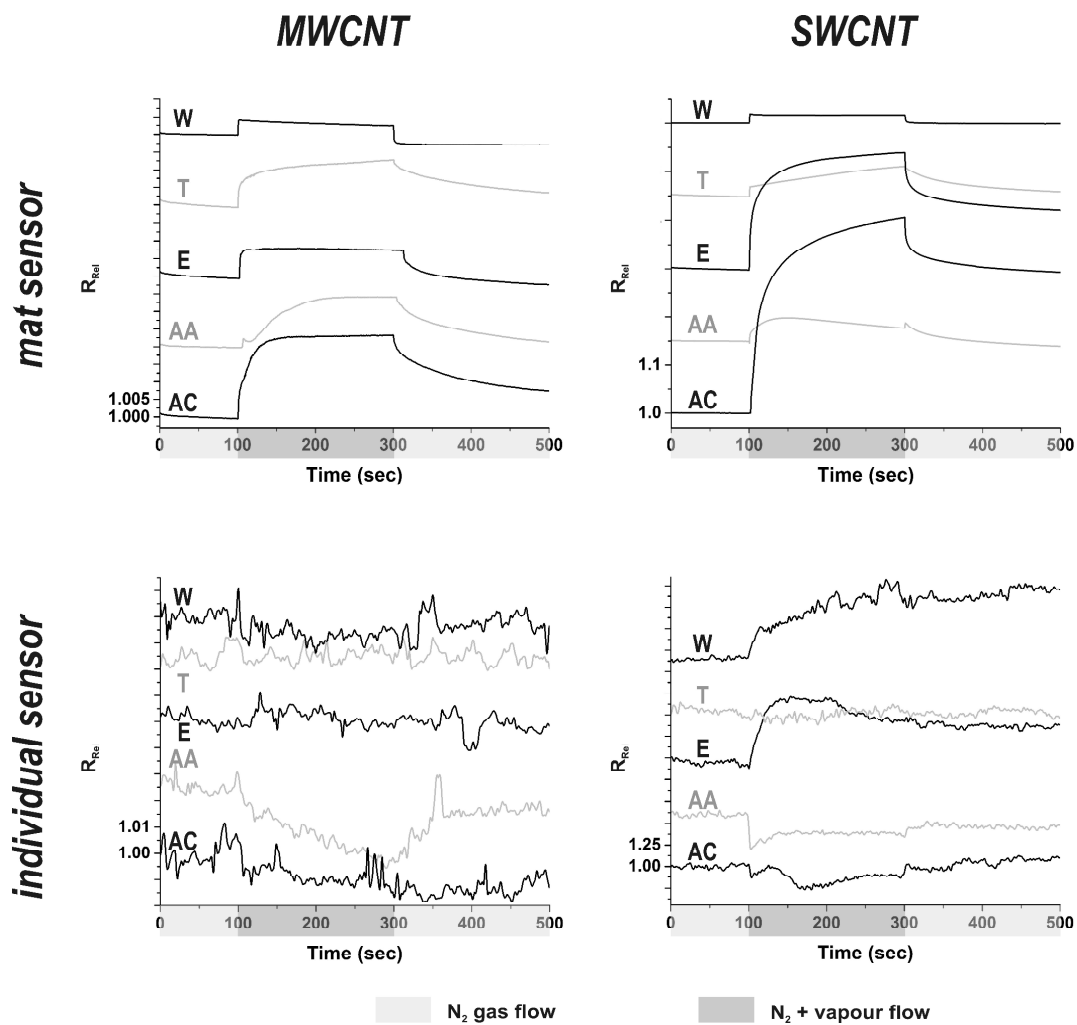

**Fig. 4** Response signals (relative resistance change) of the investigated sensors as an effect of exposition to acetone (AC), acetic acid (AA), ethanol (E), toluene (T), and water (W). Black and grey lines are showed alternating shifted vertically with a fixed $\Delta$ value, $\Delta_{\text{MWCNT mat}} = 0.02$, $\Delta_{\text{MWCNT ind.}} = 0.025$, $\Delta_{\text{SWCNT mat}} = 0.15$, and $\Delta_{\text{SWCNT ind.}} = 0.6$, to guide the eyes.

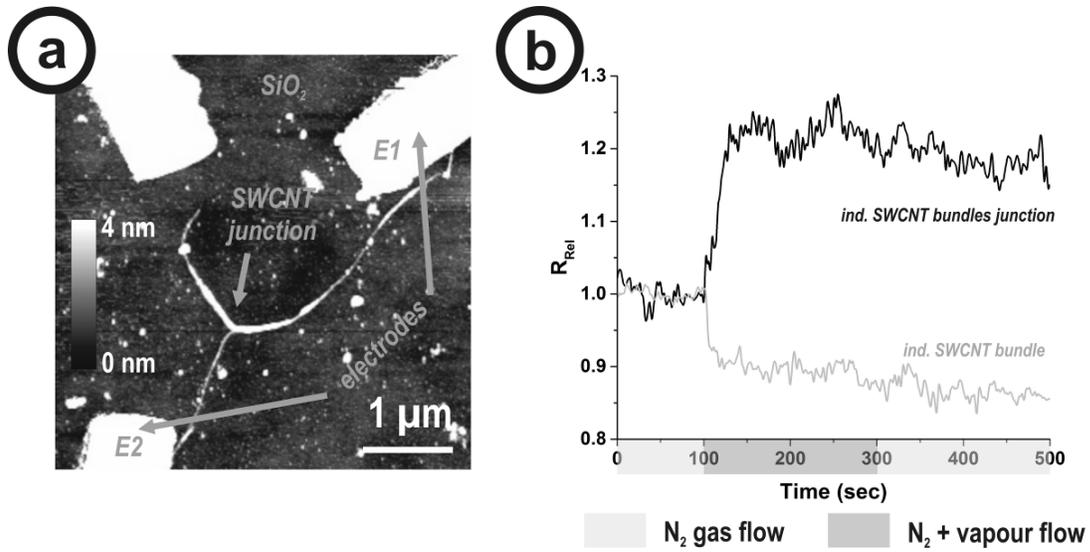

**Fig. 5** (a) AFM image of a junction of individual SWCNT bundles contacted with Cr/Au electrodes on Si/SiO$_2$ substrate; the transport measurements were taken between E1 and E2 electrodes, (b) response signals (relative resistance change) for water vapour of the above individual junction and a single wire SWCNT bundle sensor prepared on the same chip, measured simultaneously in the same nitrogen/vapour/nitrogen cycle.